\newif\ifanonymousversion
\anonymousversiontrue

\documentclass[10pt,conference]{IEEEtran}
\ifanonymousversion
\else
\IEEEoverridecommandlockouts
\fi

\usepackage{cite}
\usepackage{amsmath,amssymb,amsfonts}
\usepackage{algorithmic}
\usepackage{graphicx}
\usepackage{textcomp}
\usepackage{xcolor}
\usepackage{braket}
\usepackage{soul}
\usepackage{physics}
\usepackage[keeplastbox]{flushend}

\usepackage{tikz}
\newcommand*\circled[1]{\tikz[baseline=(char.base)]{
            \node[shape=circle,draw,inner sep=0.5pt] (char) {#1};}}

\def\BibTeX{{\rm B\kern-.05em{\sc i\kern-.025em b}\kern-.08em
    T\kern-.1667em\lower.7ex\hbox{E}\kern-.125emX}}
\begin{document}

\title{Fast Fingerprinting of\\ Cloud-based NISQ Quantum Computers}

\author{Author MacArthur}

\author{\IEEEauthorblockN{Kaitlin N. Smith\IEEEauthorrefmark{1}\textsuperscript{\textsection}, Joshua Viszlai\IEEEauthorrefmark{2}, Lennart Maximilian Seifert\IEEEauthorrefmark{2}, Jonathan M. Baker\IEEEauthorrefmark{3}\\ Jakub Szefer\IEEEauthorrefmark{6} and Frederic T. Chong\IEEEauthorrefmark{1}\IEEEauthorrefmark{2}}
\IEEEauthorblockA{\textit{\IEEEauthorrefmark{1}Super.tech, a division of ColdQuanta}\\ \textit{\IEEEauthorrefmark{2}Dept. of Computer Science, University of Chicago} \\ 
\textit{\IEEEauthorrefmark{3}Duke Quantum Center, Duke University} \\
\textit{\IEEEauthorrefmark{6}Dept. of Electrical Engineering, Yale University} \\
}
}

\maketitle
\begingroup\renewcommand\thefootnote{\textsection}
\footnotetext{Email correspondence: kaitlin.smith@coldquanta.com}
\endgroup
\thispagestyle{plain}
\pagestyle{plain}

\begin{abstract}

Cloud-based quantum computers have become a reality with a number of companies allowing for cloud-based access to their machines with tens to $>100$ qubits. With easy access to quantum computers, quantum information processing will potentially revolutionize computation, and superconducting transmon-based quantum computers are among some of the more promising devices available. Cloud service providers today host a variety of these and other prototype quantum computers with highly diverse device properties, sizes, and performances. The variation that exists in today's quantum computers, even among those of the same underlying hardware, motivate the study of how one device can be clearly differentiated and identified from the next. As a case study, this work focuses on the properties of $25$ IBM superconducting, fixed-frequency transmon-based quantum computers that range in age from a few months to approximately $2.5$ years. Through the analysis of current and historical quantum computer calibration data, this work uncovers key features within the machines that can serve as basis for unique hardware fingerprint of each quantum computer. This work demonstrates a new and fast method to reliably fingerprint cloud-based quantum computers based on unique frequency characteristics of transmon qubits. Both enrollment and recall operations are very fast as fingerprint data can be generated with minimal executions on the quantum machine. The qubit frequency-based fingerprints also have excellent inter-device separation and intra-device stability.

\end{abstract}

\section{Introduction}

We are at a transformative moment in quantum computing. After significant
investment from government, industry, and academia alike, the widespread speculation surrounding quantum information processing (QIP) is evolving into overwhelming optimism -- much of the scientific community believes that quantum computers (QCs) will revolutionize computation in a matter of decades. This excitement surrounding quantum computing can be credited to the fact that machines once thought of as an elusive theoretical concept~\cite{divincenzo2000physical} are beginning to emerge in labs across the world~\cite{gambetta2017building,egan2021fault}. These prototypes consist of a variety of technologies from devices that compute using superconducting (SC) circuits to implementations that encode information within atomic particles. 

QCs and their quantum bits, or qubits, offer a novel means to solve some of today's most challenging problems through the careful application of quantum superposition, interference, and entanglement. Quantum advantage over classical computers is on the horizon, as larger quantum computers, with exponentially larger computational spaces than their classical counterparts, emerge. Anticipated applications for robust QCs include cryptography~\cite{shor1999polynomial}, big data~\cite{grover1996fast} chemistry~\cite{kandala2017hardware}, optimization~\cite{moll2018quantum}, and machine learning~\cite{biamonte2017quantum}, among others. 

\begin{figure}[t]
     \centering
         \includegraphics[width=0.8\linewidth,trim={0cm 0cm 0cm 0cm},clip]{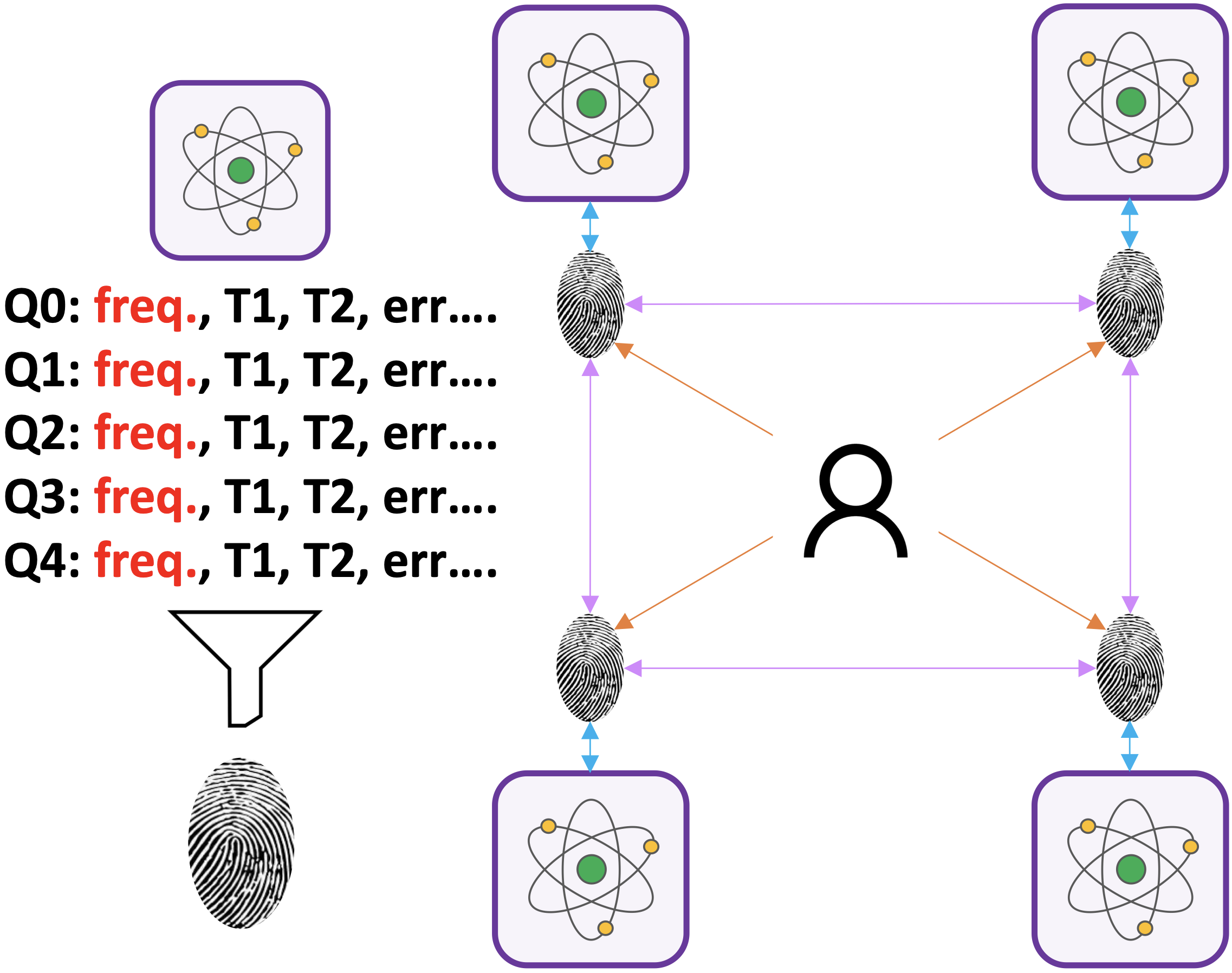}
        \caption{\small Quantum hardware fingerprints are constructed from device signatures that are robust and reproducible over time. Potential candidates for features to be used for fingerprinting include frequency of qubits, freq., decoherence characteristics, $T1$ and $T2$ times, and various error rates, err. 
        }
        \label{fig:overview}
\end{figure}

\subsection{Today's Quantum Computers}

Recent advances have allowed the development of QCs with tens to over a hundred physical qubits~\cite{eagle-127q}. The debut of QCs with a modest number of qubits marks a breakthrough in QIP, but unfortunately these devices, sometimes referred to as noisy intermediate scale quantum (NISQ) machines~\cite{preskill2018quantum}, are still plagued by high gate infidelity and short coherence windows. Regardless, the promise of quantum computing, and breakthroughs it can achieve, has led to the development of many prototype devices that compute with atoms, ions, superconducting circuits, and more. Several of these aforementioned implementations have been co-developed with quantum software development kits (SDKs) that allow for these QCs to be programmed remotely from a classical machine.

\subsection{Cloud-based Quantum Computing Today}

Quantum computing is still rapidly developing, and today there are several cloud-based QCs. Although these QCs, along with their highly-specialized, associated classical infrastructure, are in limited supply, they are readily available for remote access. 
These QCs and corresponding software stacks are primarily accessed by researchers in academia and industry, but can be used by anybody with and e-mail account and runtime credit. Current cloud vendors with their own quantum computers include industry giants like IBM, Google, Microsoft and Honeywell, as well as startups such as Xanadu, Rigetti, IonQ. Further, Amazon Braket and Microsoft Azure Quantum provide quantum computing as a service via multiple other quantum hardware vendors.

Quantum computing as a cloud service is expected to grow considerably over the next decade, continuing to be the main gateway to quantum computation targeted toward sensitive applications such as financial modeling, cryptography, and genome data analysis. The wide availability of quantum cloud devices has many benefits but has also introduced security vulnerabilities. For example, multi-tenant~\cite{8891998} and temporal~\cite{3293920} covert communication has been showcased in academia.

\subsection{Towards Hardware Fingerprinting of Quantum Computers}

One of the well-known classical hardware security primitives is a hardware fingerprint.
A hardware fingerprint includes information collected about a remote computing device for the purpose of identification. A robust fingerprint includes features that are unique, stable, and collision resistant. In classical computing, recent work fingerprints cloud FPGAs through Physical Unclonable Functions (PUFs) based on the decay of DRAM modules attached to the FPGAs~\cite{FPGA-Fingerprinting}. Separately, DRAM PUFs have been widely used to identify and authenticate DRAM chips~\cite{7951729,7284098}, or generate keys~\cite{8332526, 6423806}. Unfortunately, hardware fingerprinting can also allow a malicious party to navigate a cloud architecture, enabling attacks that survey computation scheduling, track resource allocation, or leak information~\cite{tian2020fingerprinting}.

Not unlike the fingerprinting of the classical computers in cloud computing setting, identifying device-level features that are distinct across different QCs and are reasonably stable enables hardware fingerprinting is a challenge. QCs for the foreseeable future will cloud-based resources, and QC fingerprinting could serve both offensive and defensive purposes. For example, the ability to uniquely identify QC resources in the cloud with fingerprints could enable malicious parties to monitor or spoof quantum devices and learn or disrupt sensitive quantum applications. Conversely, if a unique identifier were used within a cryptographic protocol, QC authentication would be enabled, or fingerprinting could be used to ensure reliability guarantees, by proving that user has access to indeed unique devices.
Fig.~\ref{fig:overview} shows at high-level how different properties of QCs, such as frequency of qubits, $f$, decoherence characteristics, $T1$ and $T2$ times, or various error rates, $err$, could be candidates for a source of fingerprint.

In this work, we are specifically interested in \emph{in-situ} quantum device fingerprinting i.e., the quantum device itself provides a means of identification as opposed to authentication provided by co-located classical peripheral. We are motivated to use the quantum chip itself for identification as the potential exists for the classical peripheral to be disconnected from the QC of interest and reattached to another. Similar to the classical fingerprinting discussed earlier, a fingerprint of critical importance must comprise of unique properties of a QC \emph{which have sufficient consistency}. 
An overview of quantum hardware fingerprinting is illustrated in Figure~\ref{fig:overview}.

To better understand what QC features can be applied toward a hardware fingerprint, we must understand the evolving quantum infrastructure on the cloud and, most importantly, their robust device characteristics. Here, we focus on IBM quantum systems that consist of fixed-frequency transmon qubits. There are currently over $20$ machines of various generations and sizes that are available for public experimentation via the IBM Quantum Cloud~\cite{IBMQ,IBMQS}, and the list of available QCs is constantly evolving. At first glance, many similarities among QCs such as their underlying topology and technology might make the devices seem interchangeable, however, the QCs differ in their operational characteristics and performance as a result of their hardware, control, and programming infrastructure. This provides motivation to apply QC operational characteristics and performance toward QC fingerprinting. A successful fingerprinting scheme should generalize across the spectrum of available quantum devices and device~generations.

\subsection{Paper Contributions}

To evaluate potential device characteristics that generalize across device size and generation, reliable over long periods of time, and unique to each device, we study historical data of real IBM machines and learn the property signatures unique to individual qubits. This study assimilates the properties of $25$ IBM QCs that range in age from a few months to approximately $2.5$ years. This required the collection and cleaning of $11,544$ records of calibration data, each containing information about device properties, coherence rates, fidelities, and more.

Our study reveals that qubit frequency is unique to each device and is most suitable for use as the basis for a QC fingerprint. Qubit frequency dependent on intrinsic device fabrication properties, making it very hard to duplicate exact frequencies even by the QC manufacturer. Thus, we show that qubit frequency within a QC's signature can be applied in a fingerprint due to its reliability, whereas other characteristics, such as gate error rates, cannot. We develop metric for comparing frequency based fingerprints and show excellent ability to distinguish inter- and intra-device fingerprint measurements.

\section{Threat Model}
\label{sec:motive}

This paper assumes quantum computing will be mainly a cloud-based service as cloud-based QCs are operated today by providers such as IBM~\cite{IBMQS}, Amazon Braket~\cite{amazon-braket}, and Rigetti Quantum Cloud Services (QCS)~\cite{rigetti-cloud-services}. Currently, cloud-based quantum services operate on the principle that users target a specific machine for execution of their program or circuit (which runs on a QC). The QC is usually selected due to specific hardware performance requirements (i.e. coherence times, error thresholds, number of qubits, etc.) required for a certain computational task. This is analogous to infrastructure-as-a-service model (IaaS) used by classical cloud service providers~\cite{lenk2011you}. We assume that a QC is effectively ``rented'' for a job. As a direct result of requesting specific hardware, users must often wait in a queue until their job is serviced by their selected QC. In the future, it is not infeasible that the user will have less control on what quantum hardware their job runs if multiple QCs satisfy the minimum QC performance requirements. In this scheme, known as server-less computing~\cite{baldini2017serverless}, the end user has minimal control over the underlying infrastructure used. In a server-less computing environment, workloads ideally run homogeneously from one partition to the next as providers attempt to hide intimate details of the underlying hardware from the user. This prevents users from gleaning details of the cloud resources or tracking hardware scheduling and utilization.

In our work, we assume remote, cloud-based QCs based on fixed-frequency transmon qubits, basing our device modeling on the IBM QCs featured in Table~\ref{tab:ibm-qcs}. The goal of the attacker is to infer the identity of a QC based on intrinsic characteristics directly tied to hardware. In particular, the attacker creates a fingerprint with carefully selected features from a QC’s unique property signature that includes information about coherence time, gate fidelity, and error rates, among others. While we focus our study on fixed-frequency transmon QCs, we hypothesize that the presented techniques will be applicable to alternative quantum computing platforms. We assume that the attacker has privileges to submit quantum circuits written at either the gate or pulse abstraction level in order to develop their QC fingerprint. Once a malicious party can reliably locate compute resources within the cloud, offensive behavior such as such as covert and side channel attacks, information leakage, and interference with data processing are facilitated. The victim is the benevolent QC user that accesses expensive cloud resources for the quantum acceleration of sensitive applications proposed for quantum such as cybersecurity, finance, drug design, and weather forecasting~\cite{EETimes}.

\begin{figure*}[t]
\centering
\includegraphics[width=0.9\linewidth,trim={0cm 7cm 0cm 7cm},clip]{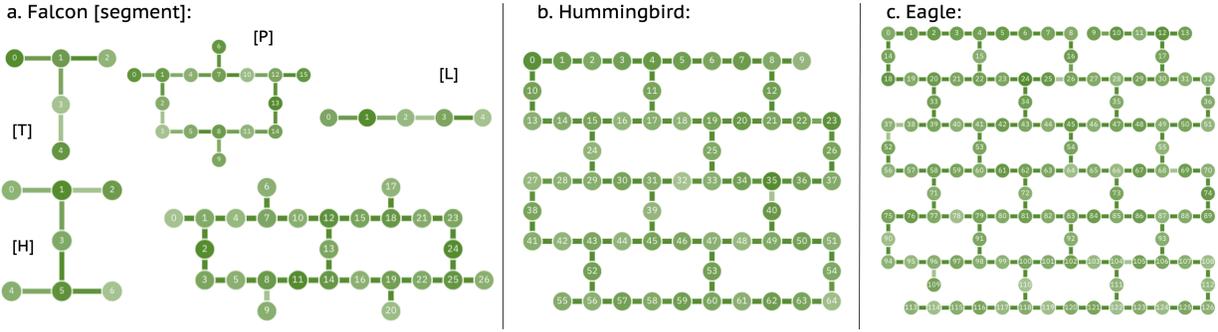}
\caption{\small IBM device topology of the (a) Falcon, (b) Hummingbird, and (c) Eagle processors. Device [segment] is indicated, where applicable. Canary processor topology has not been included since it contains a single qubit.} 
\label{fig:ibm-device-topology}
\end{figure*}

\section{Quantum computing fundamentals}
\label{background}

This section provides brief introduction to quantum computing ideas.

\subsection{Quantum Information}
\label{information-qc}

QCs use quantum bits, or qubits, that have two basis states represented as  $\Ket{0} =  \begin{bmatrix}  1 & 0
\end{bmatrix}^\intercal$ and $\Ket{1} =  \begin{bmatrix}  0 & 1
\end{bmatrix}^\intercal$. Classical bits hold a static value of either $0$ or $1$, but qubits can demonstrate states of superposition in the form of $\alpha\ket{0} + \beta\ket{1}$. Here, the probability amplitudes are complex values, $\alpha,\beta \in \mathbb{C}$, that satisfy $|\alpha|^2+|\beta|^2=1$. Superposition enables $n$ qubits to represent up to $2^n$ states simultaneously. This phenomenon, along with the ability for quantum states to interfere and become entangled, allow select problems to be solved with significant reductions in complexity in terms of compute resources or time.
Some common single-qubit transformations include the $R_x(\pi)=X$, $R_z(\pi)=Z$, and $R_y(\pi)=Y$, where $\theta = \pi$ rotation operations
cause a basis-flip, a phase-flip, and a combination basis-flip and phase-flip, respectively. These are typically the single qubit ``gates'' that users can use in their quantum circuits. 
Examples of multi-qubit operations include the logical $SWAP$ operation that causes the exchange of quantum state along with controlled gates, such as $CX$ or $CZ$, that execute an operation on a target qubit depending on the state of one (or more) control qubit(s). 
The set of basis gates provided by a particular QC is technology dependent as hardware has certain operators that are preferred for physical implementation. The basis gates usually consist of a set of single-qubit operations that are capable of implementing arbitrary rotations, $R_x$, $R_y$, and $R_z$, along with an additional two-qubit operation such as $CX$ or $CZ$. 

\subsection{Today's Physical Qubits}
\label{NISQ-info}

Current quantum devices are extremely fragile. As a result, some of the biggest challenges that limit scalability include decoherence, gate errors, readout errors and crosstalk. Additionally, most NISQ devices suffer from limited qubit-qubit communication since architecturally constrained hardware only supports nearest neighbor connectivity.

Coherence times are defined as how long a qubit maintains its state, after which its information is lost. These are $T_1$ and $T_2$ times corresponding to bit and phase information, respectively. Decoherence can present itself in the form of a qubit in a high-energy state decaying to a low-energy state or a qubit interacting with its surrounding environment. For superconducting quantum computers, coherence times have improved from one nanosecond to $100$ microseconds in the last decade. 
However, imprecise control of QCs can lead to gate and readout errors, which corrupt the intended state during gate operations and measurement, respectively. From public IBM information, single-qubit instruction error rates are of the order of $10^{-3}$, whereas for two-qubit instructions, such as $CX$, are $10^{-2}$~\cite{Tannu:2019a}. Crosstalk arises from unwanted interactions between the qubits and from leakage of the control signals onto nearby qubits which are not part of the intended gate operation but operate at similar frequencies~\cite{murali2020software}. 

Many qubit technologies based on atoms, ions, photons, and superconductors have emerged. Of interest to this study are fixed-frequency transmons, a type of SC circuit qubit realization. Fixed-frequency means that each transmon qubit's operating frequency is fixed at fabrication and does not change, as opposed to tunable-frequency transmon qubits. Transmon qubits and their fabrication are further described in Section~\ref{sec:Transmon}.

\section{Transmon-based QCs: Fabrication \& Variation}
\label{sec:Transmon}

This section presents background on transmon QC fabrication as well as various sources of static and dynamic variation among qubits in QCs.

\subsection{Device Fabrication}
\label{sec:trans-dev}

Transmons provide a viable platform for realizing physical qubits~\cite{majer2007coupling}. Transmon qubits employ Josephson Junctions (JJs) along with other hardware at ultra-cold temperatures to act as mesoscopic-scale artificial atoms with an anharmonic energy spectrum. Compared to alternative technologies, transmons are extremely promising due to recent improvements in device coherence, operation fidelity, and addressability~\cite{jurcevic2021demonstration,kjaergaard2020superconducting}. 
Progress in transmon technology has enabled early demonstrations of quantum algorithms~\cite{kandala2017hardware,o2016scalable}, quantum error correction~\cite{reed2012realization,ofek2016extending,hu2019quantum}, and quantum primacy~\cite{arute2019quantum}. 

Transmon QCs are promising for qubit scaling because their fabrication takes advantage of well-established techniques that are streamlined by decades of classical device manufacturing.
During processing, qubits are constructed in layers~\cite{nersisyan2019manufacturing,place2021new,vrajitoarea2020strongly}, and just like in classical fabrication, some degree of device variation is injected. Inevitable QC variation is due to process imprecision or stochastic defects related to the manufacturing environment or materials. 

While every SC quantum device lab may have a unique transmon qubit recipe, at a high level, qubit fabrication involves: \circled{a}\ Substrate cleaning and preparation, \circled{b}\ Base metal placement, \circled{c}\ Lithography-based feature definition, \circled{d}\ JJ definition, creation and placement, and \circled{e}\ Packaging and final test~\cite{nersisyan2019manufacturing,place2021new,vrajitoarea2020strongly}. Each of these fundamental steps presents an opportunity to introduce inter-chip and intra-chip quantum computer feature variation.

However, many SC qubit manufacturers implement custom procedures that attempt to reduce heterogeneity and boost qubit quality. For instance, prior work~\cite{place2021new} demonstrates improvements in qubit relaxation and coherence times by replacing the often-used niobium with tantalum in the initial metal layer of the chip. Other research \cite{nersisyan2019manufacturing} found that removing a thin layer of the substrate as a pre-fabrication step helps decrease qubit loss. These examples show that the route to fault-tolerant (FT) quantum computing with SC qubits necessitates refined materials and methodologies for producing QCs. Unfortunately, improved fabrication that allows QCs to scale are still likely to fail at creating perfectly homogeneous systems. Unavoidable variability in QC fabrication, even if seemingly small, elicits unique differences between devices, opening doors for robust fingerprinting of QCs. 

\subsection{Qubit Variation}
\label{sec:qc-var}

QCs are fabricated in layers, and there are many steps in processing where a slight deviation from the target specification impacts physical qubit features. Physical qubit variation causes performance variation that is  observable across a single QC and between QCs. 
QC variation is often the result of
fabrication imprecision, and physical abnormalities in a QC are referred to as defects. QC defects can be challenging to pinpoint and characterize since their impact can vary. For instance, a defect can have a subtle effect on qubit performance by slightly diminishing fidelity or coherence. At the other extreme, a defect can cause severe reliability issues or, in the worst case, can prevent qubits from participating in meaningful computation.
Variation can broadly be categorized into two forms -- static and dynamic.

\textbf{\emph{Static variation}} can occur across devices (inter-) as well as within devices (intra-), but the corresponding device characteristics are fairly stable over time.
Fabrication imprecision~\cite{hertzberg2021laser,kreikebaum2020improving} is a prime example of static variation.
JJs have incredibly small feature sizes that are hundreds of nanometers in scale~\cite{hertzberg2021laser}. Thus, small imperfections that appear in JJ positioning, component dimension, or surrounding layers influence operational characteristics of the transmon~\cite{kreikebaum2020improving}. 
In turn, variation in individual qubit characteristics influences how well qubits can work together during computation. 

Current QC fabrication techniques lack the precision required to produce devices that exactly match targeted design specifications.
A major effect of static variation is that the post-fabrication frequencies of transmon QCs are unique to each qubit and across each device.
By the nature of static variation, the uniqueness of these frequencies are fairly stable over the entire life cycle of the device.

\textbf{\emph{Dynamic variation}}, as its name alludes, fluctuates over time inter- and intra-QC.
An important example of dynamic variation is a two-level system (TLS)~\cite{muller2019towards}.
While TLS stems from fabrication imperfection, its effect transiently and stochastically appears across a quantum chip. TLS is caused by impurities inside materials or irregularities within atomic crystalline lattice structures that can appear unexpectedly in oxide layers or on the QC's surface.
TLS parasitically couples to qubits in an unpredictable manner if it appears within close proximity to the critical components of transmons, or the JJs.
When TLS appears close to, and is near-resonant with, active transmon elements, the TLS interferes with qubit activity. Qubit energy is absorbed, and the qubit experiences significant reductions in coherence that is observable in truncated $T_1$ and $T_2$ times. Coupling strength to TLS varies over time, causing time-dependent qubit property fluctuation~\cite{burnett2019decoherence,schlor2019correlating}.
Apart from TLS, other dynamic variations are observed from thermal fluctuations, magnetic flux, and quasi-particles~\cite{Klimov_2018,burnett2019decoherence}.

\textbf{\emph{Variation effects on fingerprinting:}} Differentiating variations as static and dynamic is particularly important to fingerprinting.
While all variations produce unique qubit and device signatures, robust fingerprints require the use of device characteristics stemming from static variation so that fingerprints are reproducible over time.
It should be noted that even features thought to be predominantly affected by static variation, like qubit frequency, are not entirely free from dynamic effects. Even the most reliable device features tend to show fine-grained fluctuations of fairly small magnitudes as well as occasional, coarse-grained spikes of larger magnitudes.
Thus, effective fingerprinting schemes must be robust to constant, fine-grained feature fluctuations and tolerant to occasional, coarse-grained spikes.

\begin{table*}[t]
    \centering
    \caption{\small IBM quantum computers considered in this study. * indicates unavailable feature,  $\dagger$ indicates retired device, $\string^$ indicates restricted device (7/12/2022).}
    \begin{tabular}{c|c|c|c|c|c|c}
        \textbf{Machine Name} & \textbf{Processor Type} & \textbf{Qubits} & \textbf{Quantum Volume}& \textbf{CLOPS} & \textbf{Date Online} & \textbf{Total Records}  \\
        \hline
        Armonk$\dagger$ & Canary r1.2 & 1 & 1 & * & Oct. 16, 2019 & 941\\
        \hline
        Bogota$\dagger$ & Falcon r4L & 5 & * & * & June 3, 2020 &  713\\
        Santiago$\dagger$ & Falcon r4L & 5 & * & * & June 3, 2020 & 718\\
        Manila & Falcon r5.11L & 5 & 32 & 2.8K & April 28, 2021 & 426 \\
        \hline
        Lima & Falcon r4T & 5 & 8 & 2.7K & Jan. 8, 2021 & 518\\
        Belem & Falcon r4T & 5 & 16 & 2.5K & Jan. 8, 2021 & 510\\
        Quito & Falcon r4T & 5 & 16 & 2.5K & Jan. 8, 2021 & 519\\
        \hline
        Casablanca$\dagger$ & Falcon r4H & 7 & * & * & Aug. 7, 2020 & 511 \\
        Jakarta & Falcon r5.11H & 7& 16 & 2.4K & April 28, 2021 & 440\\
        Lagos$\string^$ & Falcon r5.11H & 7 & 32 & 2.7K & May 20, 2021 & 293\\
        Perth$\string^$ & Falcon r5.11H & 7 & 32 & 2.9K & July 22, 2021 & 245\\
        Nairobi & Falcon r5.11H & 7& 32 & 2.6K & May 20, 2021 & 399\\
        Oslo & Falcon r5.11H & 7& 32 & 2.6K & March 25, 2022 & 55\\
        \hline
        Guadalupe & Falcon r4P & 16 & 32 & 2.4K & Jan. 8 2021 & 490\\
        \hline
        Montreal & Falcon r4 & 27 & 128 & 2K & June 3, 2020 &  748\\
        Toronto & Falcon r4 & 27 & 32 & 1.8K & June 3, 2020 & 743\\
        Sydney$\dagger$ & Falcon r4& 27 & * & * & Sept. 2, 2020 & 446\\
        Mumbai & Falcon r5.1 & 27 & 128 & 1.8K & Nov. 13 2020 &  572\\
        Hanoi & Falcon r5.11 & 27 & 64 & 2.3K & April 24, 2021 & 407\\
        Cairo & Falcon r5.11 & 27 & 64 & 2.4K & May 5, 2021 & 357\\
        Auckland & Falcon r5.11 & 27 & 64 & 2.4K & July 22, 2021 & 238\\
        Kolkata & Falcon r5.11 & 27 & 128 & 2K &  Nov. 13, 2020  &520\\
        Geneva & Falcon r5.8 & 27 & 32 & 1.9K & March 25, 2022 & 56\\
        \hline
        Brooklyn$\dagger$ & Hummingbird r2 & 65 & * & * & March 4, 2021 &404 \\
        \hline
        Washington & Eagle r1 & 127 & 64 & 850 & Sept. 16, 2021 &  275\\
    \end{tabular}
    \label{tab:ibm-qcs}
\end{table*}

\section{Property Analysis of the IBM QCs}
\label{sec:properties}

To develop and test the new QC fingerprinting methods, this work focuses on IBM quantum computers which are accessible to researchers and the general public.

\subsection{The IBM Quantum Ecosystem}

Fixed-frequency transmon QCs are thought to be especially favorable for scaling as they are characterized by coherence windows that extend to tens of microseconds~\cite{houck2008controlling,chang2013improved} along with gate errors approaching the fraction-of-a-percent thresholds required for fault tolerance~\cite{sheldon2016procedure}. As a result, IBM has dedicated much research to the development of transmon-based SC QCs, and since 2016, the company has allowed their prototype devices to be used by the public via the IBM Quantum Cloud services~\cite{IBMQ}. 
There are currently over 20 machines of various generations and sizes available for experimentation~\cite{IBMQS}, and from 2016 to 2022, IBM QCs have increased from 5 to 127 qubits in size~\cite{heavy-hex-blog}. As milestones are reached on IBM's quantum scaling roadmap~\cite{roadmap-IBM},
new QCs that feature cutting-edge technology are released to the quantum cloud while legacy devices are either updated or retired. Active devices can be found on the services page of IBM Quantum website~\cite{IBMQS}. Select information about retired QCs is also available~\cite{ibm-retired,qiskit-backend-info-archived}. As a note, at time of writing, all multi-qubit QCs demonstrate heavy-hex connectivity in anticipation for the eventual adaptation of the hybrid surface/Bacon-Shor error correcting code~\cite{heavy-hex-blog,chamberland2020topological}. Additionally, the IBM transmon hardware natively supports the basis gate set that includes the single-qubit $\sqrt{X}$, $X$, and $R_z(\theta)$ operations along with the two-qubit $CX$ gate.

\subsection{Investigating the IBM QCs}

QC calibration data is valuable for optimizing the use of quantum hardware. 
Through the analysis of machine properties, especially over time, unique property signatures can be used during circuit compilation to tailor applications to each QC~\cite{murali2019noise}.
However, the study of QC properties could also reveal candidate features that could applied toward hardware fingerprinting.
With this motivation, we gathered available historical data that contained daily calibration records for the IBM quantum machines.

Historical data was gathered and analyzed for 25 IBM QCs released from 2019 to 2021. The QCs under investigation are described in Table~\ref{tab:ibm-qcs}. 
Processor type, the feature that groups the QCs, is found in column 2 of Table~\ref{tab:ibm-qcs}. Processor type describes the general hardware qualities that go into the quantum machines. Processor family, where Eagle is the newest and Canary is the most mature, refers to the chip architecture. Revisions indicated with an ``r'' are design variants within a given family. Segment, such as ``L'' or ``T,'' differentiates processors comprising different sub-sections of a larger chip. Table~\ref{tab:ibm-qcs} features many revisions and segments of the Falcon family of processors.
The multi-qubit device coupling maps are included in Fig.~\ref{fig:ibm-device-topology}. 
As a note, at time of writing, six devices, Armonk, Bogota, Santiago, Casablanca, and Sydney were retired from IBM Quantum and were no longer accessible for data retrieval. Two devices, Lagos and Perth, became restricted-access devices and were also unavailable for data retrieval and analysis.

The IBM data set was created by accessing the historical data for each IBM QC. We sampled each machine's properties at 24-hour intervals, collecting all available data up to the QC's debut, or date online, on the IBM Quantum cloud services. The choice for this sampling interval is that IBM systems are typically calibrated once in a 24 hour period~\cite{sys-calibrate}.  The total number of records gathered for each QC can be found in the rightmost column of Table~\ref{tab:ibm-qcs}. The complete data set sums to 11,544 records of QC calibration data. Each record contains a wealth of information that provides insight to the QC's operational characteristics and performance. This data combined forms a QC's property signature as much of this calibration data is different for every QC and changes each cycle, showcasing the machine variation described in Section~\ref{sec:qc-var}. 
QC properties include information about qubit frequency, $T_1$, $T_2$, measurement success, operation duration, and gate error, among others. 

Raw data contains flaws such as incomplete, incorrect, or corrupted records that can skew statistical analysis if not properly managed. This was observed in the IBM data in the form of 1) records that were missing data, such as a coherence time or gate error for a single qubit, and 2) records that had incorrect data, such as a $CX$ error reported for two qubits that are not connected according to the device coupling map. Thus, we developed a methodology for cleaning the QC property data set in an attempt to draw conclusions about properties eligible for quantum hardware fingerprinting.  
Our solution for data cleaning followed the below steps:

\begin{itemize}
    \item Remove all duplicate QC cycles. This was observed when a calibration cycle surpassed the 24-hour sampling period. A possible explanation for this would be periods where the QC was offline or under maintenance.
    \item Remove invalid data or incorrect formatting. 
    An example of this case was when a calibration cycle returned an error value of 1 for all QC $CX$ operations. It was assumed that this was a day where the machine was inoperable.
    \item Remove all incomplete or incorrect records. Inconsistency such as this in a QC property data set was seen when qubits were missing attributes or and when $CX$ error was reported between qubits not coupled on the QC device. 
\end{itemize}
 
What resulted was a subset of clean and filtered IBM QC property records that could be used for additional analysis for features that are unique and reliable. 

\begin{figure*}[t]
     \centering
         \includegraphics[width=0.9\linewidth,trim={0cm 3.75cm 0cm 4cm},clip]{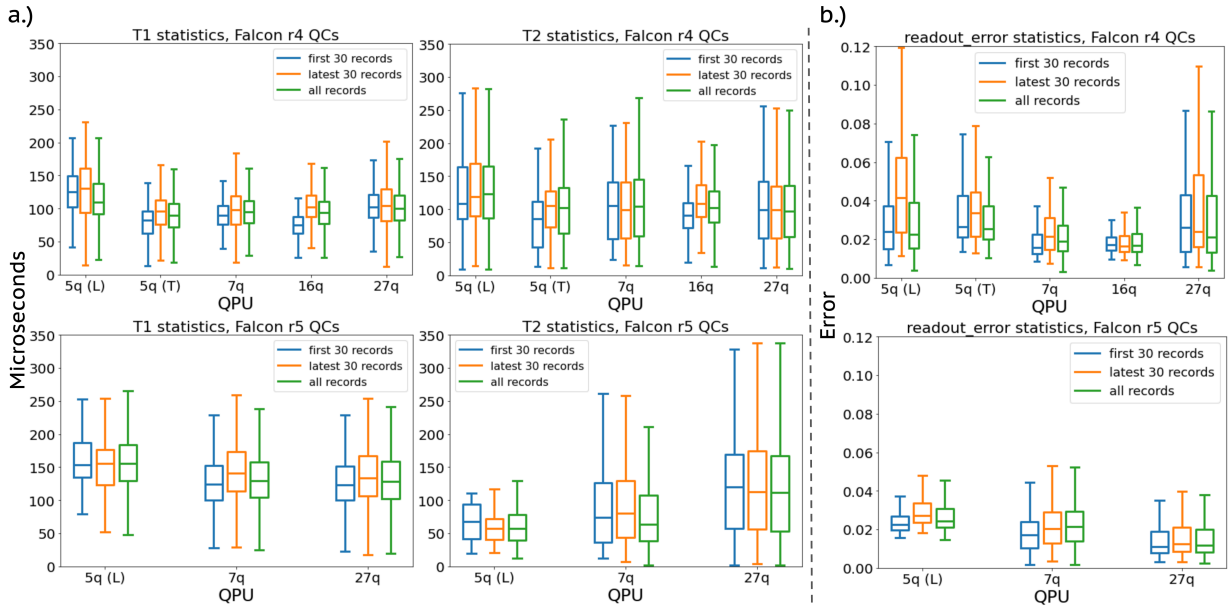}
        \caption{\small Comparison of (a) coherence properties and (b) readout error for the Falcon r4 and r5 processors. Box-and-whisker plots show distributions for the first 30 records, the latest 30 records, and all records combined. Bars are grouped by processor size in qubits and (segment).}
        \label{fig:all-falcon-compare-coherence-readout}

\end{figure*}

\begin{figure*}[t]
     \centering
         \includegraphics[width=0.65\linewidth,trim={0cm 5.6cm 0cm 5.8cm},clip]{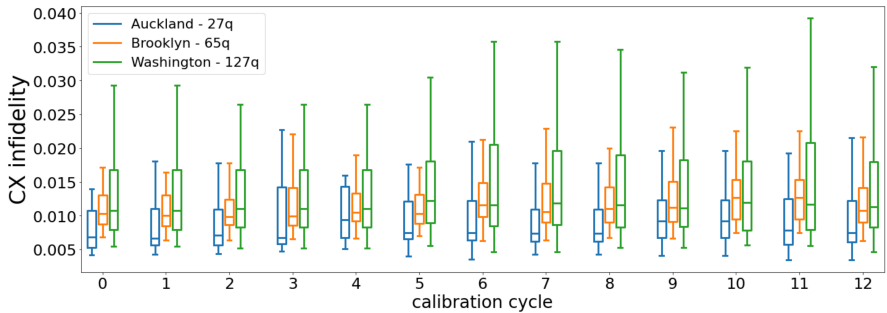}
        \caption{\small Comparison of $CX$ error statistics for three different types of IBM QC processors: Auckland (27-qubit Falcon), Brooklyn (65-qubit Hummingbird), and Washington (127-qubit Eagle). Plot includes the CX error distributions for the $12$ latest calibration cycle records.}
        \label{fig:all-cx-compare}
\end{figure*}

\subsection{Observations Between Processor Revisions}

IBM QC processors experience periodic upgrades that reflect recent advances in quantum hardware. Updates to a family of processors, such as the Falcon devices, result in a new revision~\cite{IBMQ-processors}.
From the QCs included in this study, we defined two Falcon categories, r4 and r5, and compared the properties. Devices were grouped according to size and segment. We are motivated to discover identifying features for fingerprinting that transcend hardware revisions. As an initial investigation, the QC metrics chosen for this case study were qubit coherence time, $T_1$ and $T_2$, and readout error. It should be noted that the r5 category of devices does not contain either a five-qubit T-segment device or a 16-qubit P-segment device.

Statistics about $T_1$ and $T_2$ in the form of box-and-whisker plots for each type of Falcon processor in the r4 and r5 category are described in Fig.~\ref{fig:all-falcon-compare-coherence-readout}(a). Plots contain three bars in each group, early-life distribution, latest distribution, and distribution over all records, to provide insight to performance at different periods of time as well as over all time.  
Higher is better when considering qubit coherence. When examining the plots in  Fig.~\ref{fig:all-falcon-compare-coherence-readout}(a), we once again see a large amount of variation in coherence times, not only within processor families but also over time. This finding shows the transitive nature of coherence time and suggests that this feature is not ideal for device fingerprinting. Thus, we search for alternative QC properties.

Next, we consider readout error. Qubit readout error is an important QC property to consider because of its ability to distort the final outcome of quantum computation. Lower readout error leads to more reliable QC results. Details about Falcon processor readout error are found in Fig.~\ref{fig:all-falcon-compare-coherence-readout}(b).
There are clear differences between the r4 and r5 processors when focused on readout error -- the improvements are more dramatic than those for $T_1$ and $T_2$. With both types of processors, however, we still see a large amount of overlap with error values and variation over time. As a result, we continue to search for more desirable QC fingerprint features.

\subsection{Observations Between Processor Generations}

Fig.~\ref{fig:all-cx-compare} provides statistics for $CX$ infidelity over time for three IBM processor generations. In general, the median values are similar for all processors with the Falcon device, 27-qubit Auckland, having the lowest $CX$ error. A possible explanation for this is the fact that Auckland is a more mature revision than 65-qubit Brooklyn and 127-qubit Washington, allowing more time for the technology to be refined. Another important observation of Fig.~\ref{fig:all-cx-compare} is that $CX$ error correlates with chip size -- the larger devices demonstrate higher $CX$ error upper bounds over the 12 calibration cycles. Additionally, the $CX$ error of Washington generally shows a larger distributions of $CX$ over the 12 samples. 

Two important takeaways stem from Fig.~\ref{fig:all-cx-compare}. First, although a machine might seem preferable because of one feature, such as qubit capacity, another feature critical to performance, such as high $CX$ infidelity, might make that machine less desirable. This motivates the importance of carefully pairing the right QC for a QC workload because success of an algorithm could depend on whether or not it was run on the appropriate hardware. Second, $CX$ error, while unique for each featured processor, demonstrates a significant amount of day-to-day variation, indicating that the feature might not provide the most reliable means of device identification. 

\subsection{Preliminary Fingerprinting Insights}

In the context of fingerprinting,
one would naively assume that the the error characteristics of NISQ devices might be suitable for fingerprinting.
This is not the case -- the results of the two prior sections clearly indicate that device signature components such as coherence times and error rates are unsuited to use as fingerprints which conflicts with prior work~\cite{phalak2021quantum}. $T_1$, $T_2$, measurement error, and gate error temporal variance is too large
for a reproducible, unique fingerprint. 
This motivates the study of more intrinsic device characteristics for fingerprinting, such as qubit frequency -- this is discussed in Section~\ref{sec:results}.

\section{Towards Novel and Fast Fingerprinting Based on Qubit Frequencies}
\label{sec:results}
 
As described in Section~\ref{sec:properties}, some QC features related to gate infidelity, coherence, and readout error might not be ideal for fingerprinting purposes due to issues associated with overlap among different devices and inconsistency over time. Here, we explore the viability of qubit frequency as a basis for constructing a unique QC fingerprint.

\begin{figure}[t]
     \centering
         \includegraphics[width=0.9\linewidth,trim={2.3cm 1.3cm 1.6cm 1.3cm},clip]{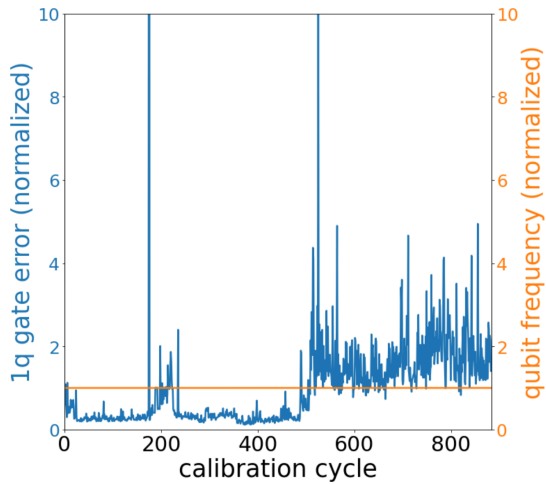}
        \caption{\small Example of features unsuitable and suitable for fingerprinting QCs: 1-qubit gate error (left axis, unsuitable) and qubit frequency (right axis, suitable) across all records (normalized by their respective mean) for IBMQ Armonk QC, a 1-qubit Canary processor.}
        \label{fig:armonk-1q-error-freq}
\end{figure}

\subsection{Feature Variation Over Time}
\label{feature-over-time}

\begin{figure}[t]
     \centering
         \includegraphics[width=0.9\linewidth,trim={0cm 1.5cm 0cm 1.9cm},clip]{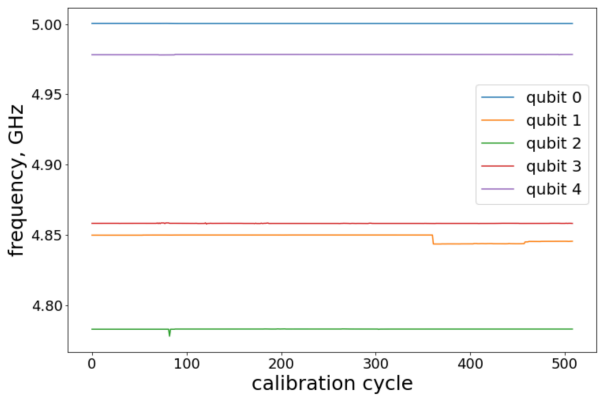}
        \caption{\small Frequency vs. calibration cycles for the five-qubit Bogota IBM QC, showing the consistency of qubit frequency over many cycles.}
        \label{fig:freq-vs-cycles}
        
\end{figure}

\begin{figure}[t]
     \centering
         \includegraphics[width=0.9\linewidth,trim={2.1cm 3.5cm 2cm 3.5cm},clip]{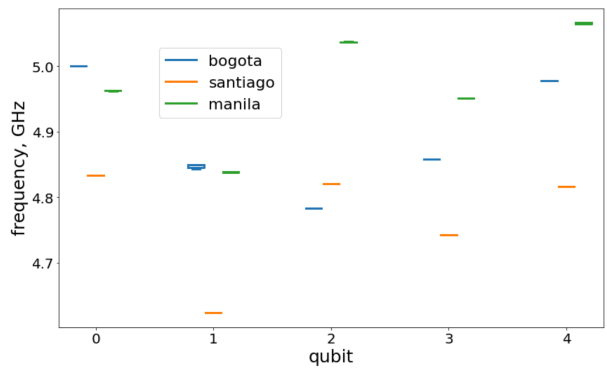}
        \caption{\small Comparison of frequency distribution vs. qubit for the five-qubit IBM QCs Bogota, Santiago and Manila.}
        \label{fig:freq-vs-qbits-b-s-m}
        
\end{figure}

Here, we compare the transient nature of single-qubit error and frequency on IBM's single-qubit Canary processor, Armonk. Fig.~\ref{fig:armonk-1q-error-freq} includes two plots. Single-qubit gate error, i.e. $X$ or $SX$ gate error, is indicated on the left axis in blue while qubit frequency is indicated on the right axis in orange. The curves describe the error and frequency data normalized by their means, respectively, for calibration cycles collected over approximately 2.5 years. When analyzing Fig.~\ref{fig:armonk-1q-error-freq}, we observe that 
the recorded values for gate error are much more noisy and have a larger degree of variation as compared to frequency.

Fig.~\ref{fig:armonk-1q-error-freq} demonstrates that qubit frequency does not fluctuate at the same extent as gate error. However, slight variation in frequency exists. As an example we analyze qubit frequency over device lifetime for the five-qubit IBM Bogota QC. Fig.~\ref{fig:freq-vs-cycles} shows frequency vs. qubits for all of Bogota's qubits -- it provides a better understanding of intra-chip frequency distribution over time. We see occasional spikes and shifts in qubit frequency, but overall, frequency for each qubit is stable. Variation for each qubit's frequency stays relatively close to the mean frequency value and significant changes are infrequent. It is important to note that all qubits also have adequate spacing between frequency values.

\subsection{Inter-QC variation}
\label{sec:diverse-machines}

Fig.~\ref{fig:freq-vs-cycles} revealed the distinguishabiliy and consistency of an individual QC's qubit frequencies over time. This motivates the exploration of the extent that QC qubit frequencies vary among devices of similar size and performance. Fig.~\ref{fig:freq-vs-qbits-b-s-m} features a box-and-whisker plot showing frequency vs. qubit for three, five-qubit IBM QCs: Bogota, Santiago, and Manila. These devices are all Falcon processors with an L-segment connectivity graph, illustrated in Fig.~\ref{fig:ibm-device-topology}. One might conclude that all of these devices have the same operational characteristics because of their cosmetic similarities, however, despite a common processor family and layout, the frequencies associated with each of the three Falcon processors varies drastically. Fig.~\ref{fig:freq-vs-qbits-b-s-m} shows that frequency values continue to be constrained within a limited range and have definite values that are adequately spaced on each device. Further, we observe that the distributions of frequencies are different device-to-device with minimal overlap between QCs. This observation motivates us to target qubit frequency as a means to uniquely and consistently identify QCs.

\subsection{Identifying Key Features for Fingerprinting}
\label{sec:fingerprinting}

We continue our analysis of QC properties that comprise of a device's entire unique signature to quantitatively discover features that are viable for QC fingerprinting. To this end, we compare frequencies, $T_1$ / $T_2$ coherence times, and readout errors between the qubits of the five-qubit devices Bogota, Manila and Santiago. Using the 100 most recent calibration records for a specific feature (i.e. frequency, $T_1$, etc.), $\vec{x}_k = \mqty[x_{k,1} & x_{k,2} & \dots & x_{k,100}]^\intercal$ of qubit $k$, we compute the scaled Euclidean distance between the records of two qubits $i$ and $j$
\begin{equation}
    d\qty(\vec{x}_i, \vec{x}_j) = \frac{\norm{\vec{x}_i - \vec{x}_j}_2}{\sqrt{100} \Delta_\text{max}}.
\end{equation}
Here, $\Delta_\text{max}$ denotes the maximum variation for this feature over all qubits' 100 calibration cycles. Distances $< 1$ means that the qubits $i$ and $j$ are similar with respect to this feature, while distances $\gg 1$ indicate substantial dissimilarity. 

\begin{figure*}[t]
    \centering
    \includegraphics[width=0.95\linewidth]{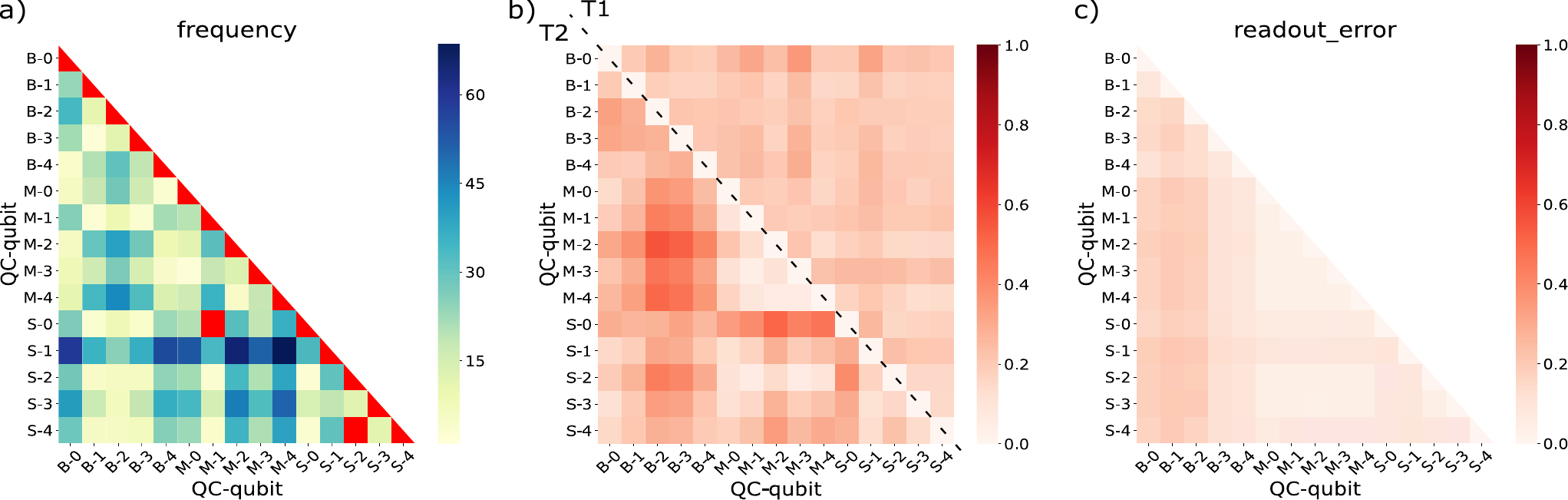}
    \caption{\small Distances of qubit features across the five-qubit processors Bogota (B), Manila (M) and Santiago (S). Values $<1$ correspond to similar attributes while values $\gg 1$ indicate dissimilarity, and in all plots darker color (higher value) means greater dissimilarity. (a) Qubits are fabricated from a wide range of frequencies, thus they generally differ a lot on the same device as well as across devices. Similar frequencies are highlighted in red. (b) $T_1$ and $T_2$ times as well as (c) readout error fluctuate across qubits but not significantly more than they fluctuate for the same qubit over time. This leads to great similarity (low distance) across devices, rendering those features less favorable for fingerprinting.}
    \label{fig:feature-triangles}
\end{figure*}

Fig. \ref{fig:feature-triangles} visualizes the feature distances for all qubits of all three machines, where the tick labels consist of the initial letters of the processors (B, M, S) and the qubit indices. In accordance with Fig. \ref{fig:freq-vs-qbits-b-s-m}, we observe great differences in qubit frequency intra- as well as inter-device as shown in Fig. \ref{fig:feature-triangles}(a). Red fields represent few instances of similarities, or collisions, between frequencies. Overall, this makes qubit frequency a strong candidate for a QC fingerprint. On the contrary, $T_1$, $T_2$, and readout errors overlap across different qubits, leading to low distances as can be seen in Fig. \ref{fig:feature-triangles}(b),(c). That means that these features do not vary more across qubits than they fluctuate over time for the same qubit, therefore they are less favorable properties for uniquely identifying QCs. This is in agreement with the distributions displayed in Fig. \ref{fig:all-falcon-compare-coherence-readout}, where we see great overlap of these features across different architectures. From this observation we conclude that the use of coherence times and readout error within a fingerprint would not be favorable as many instances of collision between identifiers would occur.

As a conclusion of this analysis, we identify frequency in fixed-frequency transmon systems as a potential resource for developing a QC fingerprint. Frequency is just one of many properties in a QC property signature that characterises a QC as a result of device variation. We investigate this special feature in more detail in the following section.

\subsection{Frequency Vector as a Fingerprint}

We define our frequency-based fingerprint of a QC as a vector of its qubit frequencies, $\vec{f} = \begin{bmatrix} f_0 & f_1 & \dots & f_{N-1} \end{bmatrix}^\intercal$, where $f_k$ is the frequency of qubit $k$. Given the frequency vectors $\vec{f}_i$ and $\vec{f}_j$ of two $N$-qubit processors $i$ and $j$, we use a normalized Hamming distance 
\begin{equation} \label{eq:J}
    \begin{aligned}
        \frac{\text{\# frequencies that differ}}{\text{\# qubits}} = \frac{\qty|\qty{k : \abs{f_{i,k} - f_{j,k}} > \Delta_{\text{avg}}}|}{N}
    \end{aligned}
\end{equation}
to estimate how similar the processors' frequency characteristics are. Here we consider two frequencies to be different if they are more than a threshold, $\Delta_{\text{avg}}$, apart.

\begin{figure}[htbp]
    \centering
    \includegraphics[width=\linewidth,trim={0cm 0cm 0cm 0cm},clip]{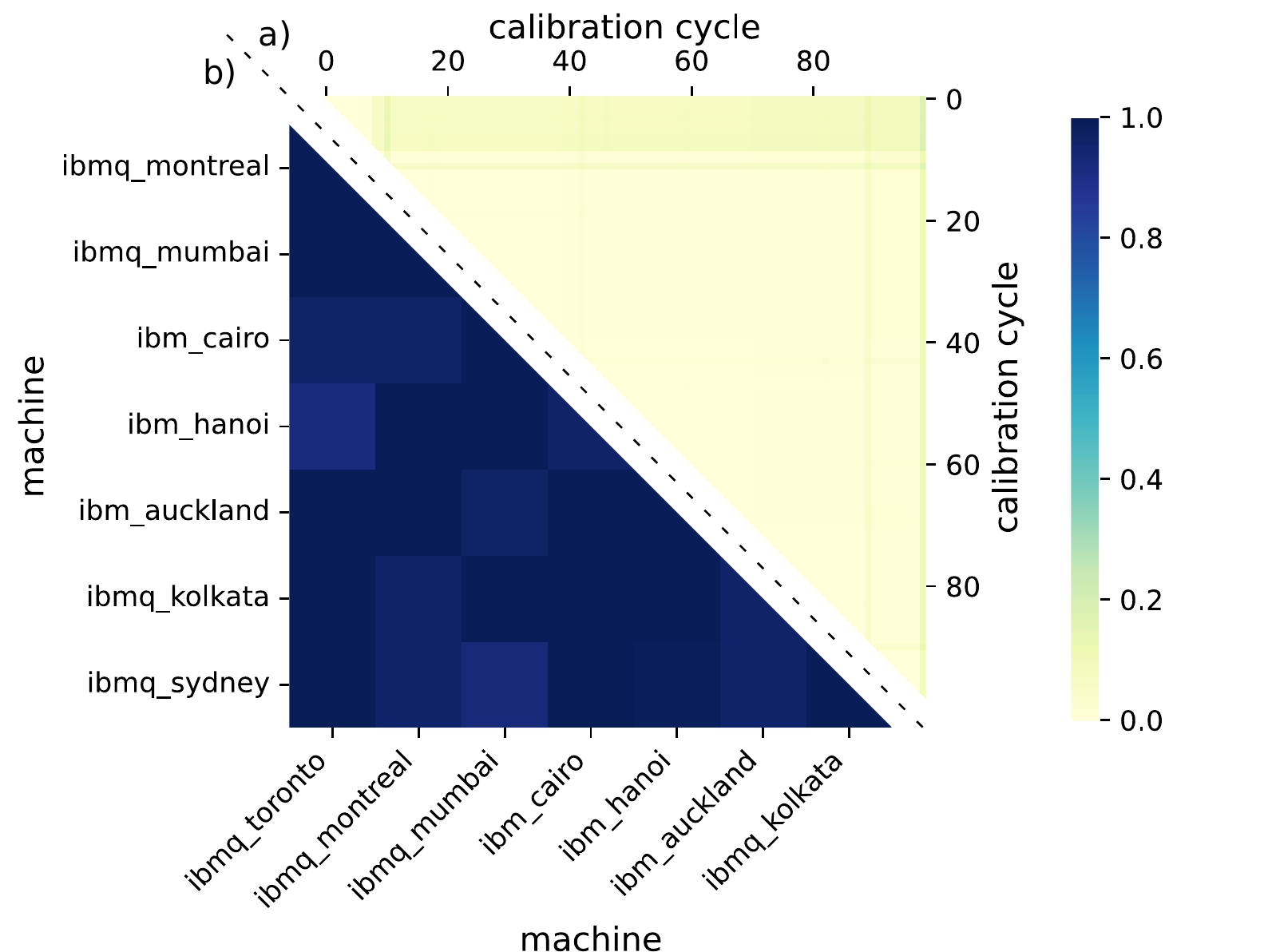}
    \caption{\small Evaluation of frequency-based fingerprints on eight 27-qubit machines over 100 calibration cycles. (a) Distances between calibration cycles for a given machine (values are averaged over all machines). The low observed distances indicate the fingerprint is stable for a given machine over time. (b) Distances between machines for a given calibration cycle (values are averaged over all cycles). The high observed distances indicate the fingerprints are unique between machines at a given time. All distances are calculated using Eq.~\eqref{eq:J}.}
    \label{fig:freq-triangles}
\end{figure}    

To demonstrate the efficacy of qubit frequencies as a unique and stable fingerprint, we evaluate frequency fingerprints from eight 27-qubit QCs over 100 calibration cycles. The QCs used represent IBM's Falcon r4 and r5 processors ranging in age from a couple months to over 2 years old. Fig.~\ref{fig:freq-triangles} shows the scaled Hamming distances between these fingerprints. The threshold, $\Delta_{\text{avg}}$, was chosen as the maximum variation in a qubit's frequency across calibration cycles averaged over all machines and qubits. 

The results in Fig.~\ref{fig:freq-triangles} support that a machine fingerprint can be constructed from qubit frequencies. The fingerprint is \textit{stable}, there is low distance between fingerprints of the same machine, and \textit{unique}, there is high distance between fingerprints of different machines.

\subsection{Considerations for a Frequency-based Fingerprint}
 
Here, we discuss considerations for a frequency-based quantum computer fingerprints. 

\textbf{\emph{Tuning of qubit frequency:}} Historical data analysis has shown the consistency of qubit frequency over time, however, there exist methods that purposefully adjust the frequency of fixed-frequency transmons. Laser annealing selectively tunes qubit frequency to bring it within a range that improves two-qubit interactions with its nearest neighbors~\cite{hertzberg2021laser,zhang2020high}, and it has been applied to improve the yield of IBM QCs with average two-qubit gate performance of $\lesssim 1\%$. Since procedures such as laser annealing enable device improvements via changing critical fingerprint components, possibly causing authentication to fail, it must be assumed that fingerprint re-enrollment is necessary, especially after device maintenance. If qubits frequencies are altered after routine QC maintenance, that device can be considered an updated version of the QC with a ``new'' intrinsic fingerprint that incorporates the frequency~changes.
Also, laser annealing the methods in~\cite{hertzberg2021laser} are imperfect, preventing annealing processes from achieving target frequencies with 100\% accuracy.

\textbf{\emph{Implications of scaling:}} As QCs are manufactured with more qubits, it is likely that the majority of their frequency distributions will be unique. However, more populated distributions could result in qubit frequencies that nearly collide, causing confusion between qubits and errors in identification. In this case, it is essential to develop protocols that include enough information (i.e. enough qubit frequency permutations) to discern one device from another, even in the case of occasional qubit frequency overlap. 

\textbf{\emph{Access to low-level control:}} Qubit frequency is determined by sweeping a qubit with discrete microwave control pulses that span a range of frequencies. The signal frequency that best resonates with the qubit is determined to be the qubit's frequency. Implementing frequency sweeps require low-level pulse control~\cite{pulse2}. Currently, the IBM family of QCs enables pulse-level control and includes qubit frequency as an element of the backend properties that can be queried by the end-user that has access to the device. Thus, qubit frequency for a QC is easily determined, making the frequency fingerprint for IBM QCs in the quantum cloud open knowledge. 
Obscuring qubit frequency to discourage device fingerprinting is possible. For example, if low-level access to QCs was restricted, forcing users to write quantum programs at gate-level abstractions, qubit frequency would be challenging to pinpoint by attackers wishing to identify a QC. Alternatively, if the cloud provider wishes to allow low-level control for their machines while concealing frequency information, an QC antivirus~\cite{deshpande2022towards} could be developed that is trained to detect and prevent frequency sweep operations that are initialized by an end-user.

\subsection{Alternative Applications of the QC Fingerprint}

Here we propose a frequency-based fingerprint based on the intrinsic properties of fixed-frequency transmon machines. Our considered threat model, Section~\ref{sec:motive}, proposed QC fingerprinting for malicious practices. However, secure cryptographic protocols along with a hardware fingerprint would enable QC authentication.
Such a scheme could facilitate the identification of expensive quantum cloud resources and the verification machine performance. Additionally, a quantum hardware fingerprint could be critical in distributed quantum applications that demand on-the-fly quantum authentication, without which any potential quantum advantage would be lost.

In this potential quantum authentication scheme, we assume that qubit frequencies are concealed by the cloud provider. Further, user access to low-level pulse control or frequency sweep experiments is restricted. When qubit frequencies are concealed, challenge-response based authentication schemes could be developed that combine permutations of qubit frequencies with a one-way function. In our analysis, frequency has shown to be both unique, reliable, and collision-resistant, making it a good candidate for use within a physically unclonable function (PUF)~\cite{maes2010physically} that identifies cloud-based quantum compute resources. We leave the complete design of this QC authentication scheme for future work.

\section{Discussion}

Past work exists that describes machine variation. For example, 
the stability of quantum machines over time is described~\cite{dasgupta2021stability}. In~\cite{dasgupta2021stability}, the authors quantify the similarity of several NISQ devices by comparing gate fidelities, duty cycles, and register addressability across temporal and spatial scales. Like our study,~\cite{dasgupta2021stability} agrees with the modest body of work that notes the variability of current quantum machines. Other prior works that discuss variation in QCs, either among individual devices or over time, include~\cite{tannu2019ensemble,murali2019noise,tannu2019not,ravi2021adaptive}. 

In the space of securing quantum computation, proposals for quantum antivirus exist~\cite{deshpande2022towards} along with quantum hardware based physically unclonable functions~\cite{phalak2021quantum}. Additionally, there is prior art in the space of quantum fingerprinting~\cite{allen2021short}. Our work differs from these past efforts by proposing a frequency-based fingerprint on the basis of uniqueness, reliability, and collision resistance discovered through historical data analysis.

\section{Conclusion}

With QCs evolving into a primarily cloud-provided commodity, the ability to fingerprint quantum hardware within the quantum cloud has serious security implications. 
In this paper, we showed how QCs based on fixed-frequency transmon qubits can already be fingerprinted. We used historical data from IBM QCs over the past 2+ years and introduced a frequency-based fingerprint that is  stable, unique, and generalizable across different device generations. We argue that these successful features emerge from the variability inherent in fabrication of fixed-frequency transmon QCs today.

\section*{Acknowledgment}

This work is funded in part by EPiQC, an NSF Expedition
in Computing, under award CCF-1730449; in part
by STAQ under award NSF Phy-1818914; in part by NSF
award 2110860; in part by the US Department of Energy Office 
of Advanced Scientific Computing Research, Accelerated 
Research for Quantum Computing Program; in part by the 
NSF Quantum Leap Challenge Institute for Hybrid Quantum Architectures and Networks (NSF Award 2016136); and in part based upon work supported by the 
U.S. Department of Energy, Office of Science, National Quantum 
Information Science Research Centers.  This research used resources of the Oak Ridge Leadership Computing Facility, which is a DOE Office of Science User Facility supported under Contract DE-AC05-00OR22725. FTC is Chief Scientist for Quantum Software at ColdQuanta and an advisor to Quantum Circuits, Inc.

\bibliographystyle{IEEEtran}
\bibliography{refs}

\end{document}